\documentclass[
reprint,
showpacs,
preprintnumbers,
nofootinbib,
amsmath,
amssymb,
aps,
prd,
]{revtex4-1}
\usepackage{enumerate}
\usepackage{color}
\usepackage{graphicx}
\usepackage{dcolumn}
\usepackage{bm}
\usepackage{empheq}
\usepackage{nomencl}
\usepackage{hyperref}
\usepackage{mathrsfs}
\usepackage{amsmath}
\hypersetup
{
	colorlinks,%
	citecolor=blue,%
	linkcolor=magenta,%
	urlcolor=blue,%
}

\usepackage{subfigure}

\begin{document}
	\newcommand{\red}[1]{\textcolor{red}{#1}}
	\newcommand{\green}[1]{\textcolor{green}{#1}}
	\newcommand{\blue}[1]{\textcolor{blue}{#1}}
	\newcommand{\cyan}[1]{\textcolor{cyan}{#1}}
	\newcommand{\purple}[1]{\textcolor{purple}{#1}}
	\newcommand{\yellowbox}[1]{\colorbox{yellow}{#1}}
	\newcommand{\purplebox}[1]{\colorbox{purple}{#1}}
	\newcommand{\yellow}[1]{\textcolor{yellow!70!red}{#1}}
	\title{Quasibound states of charged dilatonic black holes}
	\author{Yang Huang $^1$}\email{sps\_huangy@ujn.edu.cn}
	\author{Hongsheng Zhang $^{1}$}\email{sps\_zhanghs@ujn.edu.cn}
	\affiliation{
		$^1$ School of Physics and Technology, University of Jinan, 336, West Road of Nan Xinzhuang, Jinan 250022, Shandong, China}
	
	
	\begin{abstract}
	Investigation of quasibound states of black holes is significant for expected ultra light particles, as well as black holes through gravitational waves. We first investigate quasibound states of
	a massive scalar field for dilatonic charged black holes via numerical analysis. We study the complex eigen frequencies of the massive scalar field in a wide range of gravitational fine structure constant $\mu M$ in detail, and show the effects of charge. Further, we study the eigen frequencies of the massive field through analytical approach by matching the near horizon solution and far field solution, and find its spectra for excited states by iteration method. We demonstrate the numerical solution and analytical solution perfectly agree with each other in the region where charge of the black hole is large, both for real and imaginary parts of the eigen frequencies.

	\end{abstract}
	
	\maketitle
	
\section{Introduction}

Study of bound system of the sun and planets initiates classical mechanics, and study of bound state of hydrogen atom  initiates quantum mechanics. With discovery of gravitational waves and confirmation of black holes,
 bound state, scattering state, and quasinormal excitation take more and more significant status in black hole dynamics and astroparticle physics. Because of distinctive properties of black holes, generally there no real
 bound states for black hole, since any wave will infiltrate to the interior of the hole via quantum effects. The matter waves around a black hole will decay slowly or swiftly. Thus we called such state quasibound state.

 For astrophysical black hole, the mass of quasibound particle is so tiny that it is difficult to observe at colliders or accelerator in terrestrial labs. For example axion, which is supposed to explain strong CP violation in QCD, is illusive in terrestrial labs,
 since  its mass is expected to be $10^{-5}$ eV or much smaller \cite{Klaer2017,Marsh}. 
  Black hole presents increasing
  prominence to probe ultra light particles. The particles around a hole may excite more similar ones if they have proper frequencies. This type of radiation is a non-thermal one  (superradiance), which is different from Hawking radiations. At same time, the particles will leak into the black hole because of quantum tunnelling. If the two processes get detailed balance, the particles can form a ``cloud" around a hole \cite{Penrose:1971uk,Press:1972zz,Press:1973zz,Teukolsky:1974yv,Hod:2014baa,Herdeiro:2014goa}. 
   Through explorations of high resolution image of black hole filmed by Event Horizon Telescope, it expected to probe the ultra light bosonic particles based on the  birefringent effects of electromagnetic waves \cite{Chen:2019fsq}.

  Gravitational wave becomes a powerful probe to several dark astrophysical processes, which are difficult to see by traditional optical/electromagnetic observations, especially for the object which has no electromagnetic
  interaction, for example dark matters. Nothing can escape from gravity interaction. The ultra light particles surrounding black holes imprint gravitational waves of black hole binaries. Actually, even a single gravitational-wave measurement can effectively sense the existence of ultra light bosons surrounding  the gravitational wave source \cite{Hannuksela:2018izj}. 
  The observations of gravitational waves also reveal signals for bosonic cloud composed by bound state of ultra light particles through spin-induced multipole moments and tidal Love numbers \cite{Baumann:2018vus}.
  Recently, it is shown that one could find the signal of dark matters around black hole binary from gravitational tail wavelet \cite{Lin:2019qyx}.

  Quasibound states of a black hole have been studied under several different conditions \cite{Cayuso:2019ieu,Qiao:2020fta,Li:2020vid,Huang:2018qdl,Huang:2017nho,Huang:2016zoz,Huang:2016qnk,Detweiler:1980uk,Furuhashi:2004jk,Huang:2017nho,Zhou:2013dra,Barranco:2012qs}. The minimal extension of general relativity, i.e., the dilatonic gravity or called scalar tensor theory,
  can be traced back to Kluza-Klein compactification and Dirac's large number hypothesis, in which only one new degree is introduced. The quasinormal modes of charged dilatonic black hole is studied in \cite{Ferrari:2000ep,Li:2001ct}. 
  The superradiant instability and charged scalar clouds of the dilatonic black hole is studied in \cite{Huang:2017whw,Siahaan:2015xna,Bernard:2016wqo,Bernard:2017rcw,Li:2013jna,Li:2015bfa,Li:2014fna}.
  Recently, the scattering properties of such a hole is studied in \cite{Huang:2020bdf}. In this paper, we will discuss the quasibound state of a charged dilatonic black hole.

  This paper is organized as follows. In the next section, we present the theory frame and our numerical method in detail. In section III, we develop an analytical approach for the complex eigen frequency, especially 
  for a black hole with large charge. In section IV, we present our main result, and demonstrate that the results from numerical method are well consistent with results from analytical method, particularly in the case of 
  large charge limit. In section V, we concisely conclude this article.    
%

\section{Massive scalar field in the GMGHS spacetime}\label{Sec: setup}
\subsection{The background metric}
For dilatonic charged gravity, the line element of charged dilatonic black hole is found by Gibbons and Maeda \cite{Gibbons:1987ps}, and independently by Garfinkle, Horowitz, and Strominger \cite{Garfinkle:1990qj} (GMGHS) in Einstein frame,
\begin{equation}
ds^2=-Fdt^2+F^{-1}dr^2+r^2G\left(d\vartheta^2+\sin^2\vartheta d\varphi^2\right),
\end{equation}
with
\begin{equation}
F(r)=1-\frac{2M}{r},\;\text{and}\;G(r)=1-\frac{Q^2}{Mr},
\end{equation}
where $M$ and $Q$ are the mass and charge of the black hole respectively.
The Maxwell field and dilaton field read,
\begin{equation}
F_M=Q\sin\vartheta d\vartheta\wedge d\varphi,
\end{equation}
and,
\begin{equation}
e^{-2\phi}=e^{-2\phi_0}\left(1-\frac{Q^2}{Mr}\right),
\end{equation}
respectively.
$\phi_0$ denotes the value of the dilaton $\phi$ at spacelike infinity. $\phi_0=0$ implies an asymptotic flat manifold.
Event horizon of the GMGHS black hole is located at $r_+=2M$.
The area of the sphere goes to zero when $r=r_-=Q^2/M$ and the surface is singular.
For $Q<Q_{\mathrm{max}}\equiv\sqrt{2}M$, the singularity is enclosed by the event horizon.
In the extremal case $Q=Q_{\mathrm{max}}$, and the singularity coincides with the horizon.
Following \cite{Huang:2020bdf}, we introduce a normalized charge $q=Q/Q_{\mathrm{max}}$.
To better manifest the behavior of quasibound state frequencies in the near extremal limit, we also parameterize the black hole charge by
\begin{equation}\label{Eq: eta}
q=1-e^{-\eta}.
\end{equation}
Clearly, the black hole charge $q$ increases monotonically with $\eta$. The Schwarzschild black hole corresponds to $\eta=0\;(q=0)$, while the extremal GMGHS black hole corresponds to $\eta\rightarrow\infty$ $(q\rightarrow1)$.

\subsection{Massive Klein Gordon equation}\label{subSec: KG eq}
The massive Klein-Gordon equation governs a scalar field $\Phi$ of mass $\mu$ is $\nabla_\mu\nabla^\mu\Phi=\mu^2\Phi$.
Decomposing the scalar field as $\Phi=e^{-i\omega t}R_{\omega l}(r)Y_{lm}(\vartheta,\varphi)$ yields
the radial equation
\begin{equation}\label{Eq: the radial eq}
	\Delta\frac{d}{dr}\left(\Delta\frac{dR_{\omega l}}{dr}\right)+\left[G(r)^2\omega^2r^4-U\right]R_{\omega l}=0,
\end{equation}
where
\begin{equation}
	U=\Delta\left[l(l+1)+\mu^2r^2G(r)\right].
\end{equation}
It is useful to introduce a new radial function as
\begin{equation}
	\psi_{\omega l}(r)=\frac{R_{\omega l}(r)}{r\sqrt{G(r)}}.
\end{equation}
Then the radial equation becomes 
\begin{equation}\label{Eq: the radial eq psi}
	\frac{d^2}{dx^2}\psi_{\omega l}+\left[\omega^2-V_l(r)\right]\psi_{\omega l}=0,
\end{equation}
where $x=\int dr/F$ is the tortoise coordinate, and the effective potential is given by
\begin{equation}\label{Eq: effective potential}
	\begin{aligned}
	V_l(r)=&\frac{F(r)}{G(r)}\left[\frac{F'(r)}{r}+\frac{l(l+1)}{r^2}+\mu^2G(r)\right]\\&-\frac{2M^2q^2}{r^4}\frac{F(r)}{G(r)^2}\left[1+\frac{q^2}{2}\left(1-\frac{6M}{r}\right)\right].
	\end{aligned}
\end{equation}
Clearly, if we take $q=0$, then $G(r)=1$ and $V_l(r)$ reduces to the effective potential of the massive scalar field in the Schwarzschild spacetime. $q=0$ also implies that the dilaton $\phi$
 vanishes. Thus the dilaton $\phi$ is not independent hair.

At the horizon, $V_l(r)$ vanishes for $q\neq1$\footnote{We only consider the non-extremal case, i.e., $0\leq q<1$.}, so the asymptotic solution to Eq.(\ref{Eq: the radial eq psi}) is
a superposition of ingoing and outgoing waves. Regularity requires a purely ingoing wave solution
at the horizon,
\begin{equation}\label{Eq: bcs horizon}
	\psi_{\omega l}\sim e^{-i\omega x}\sim\left(r-r_+\right)^{-2iM\omega}.
\end{equation}

At infinity, the potential tends to $\mu^2$, and the radial function has the following asymptotic behavior
\begin{equation}\label{Eq: bcs infty}
\psi_{\omega l}\sim r^{\chi}e^{\rho r},
\end{equation}
where
\begin{equation}\label{Eq: def of chi and k}
	\chi=\frac{M(\mu^2-2\omega^2)}{\rho},\;\;\text{with}\;\;\rho=\pm\sqrt{\mu^2-\omega^2}.
\end{equation}
The behavior of the radial function at large distance is determined by the sign of the real part of $\rho$.
If $\mathrm{Re}(\rho)>0$, the function is divergent, whereas if $\mathrm{Re}(\rho)<0$, the function tends to zero. Here, we are interested in quasibound state solutions and choose $\mathrm{Re}(\rho)<0$. Before proceeding, it would be helpful to analyze the behavior of the effective potential.

\begin{figure*}
	\includegraphics[width=0.48\textwidth,height=0.38\textwidth]{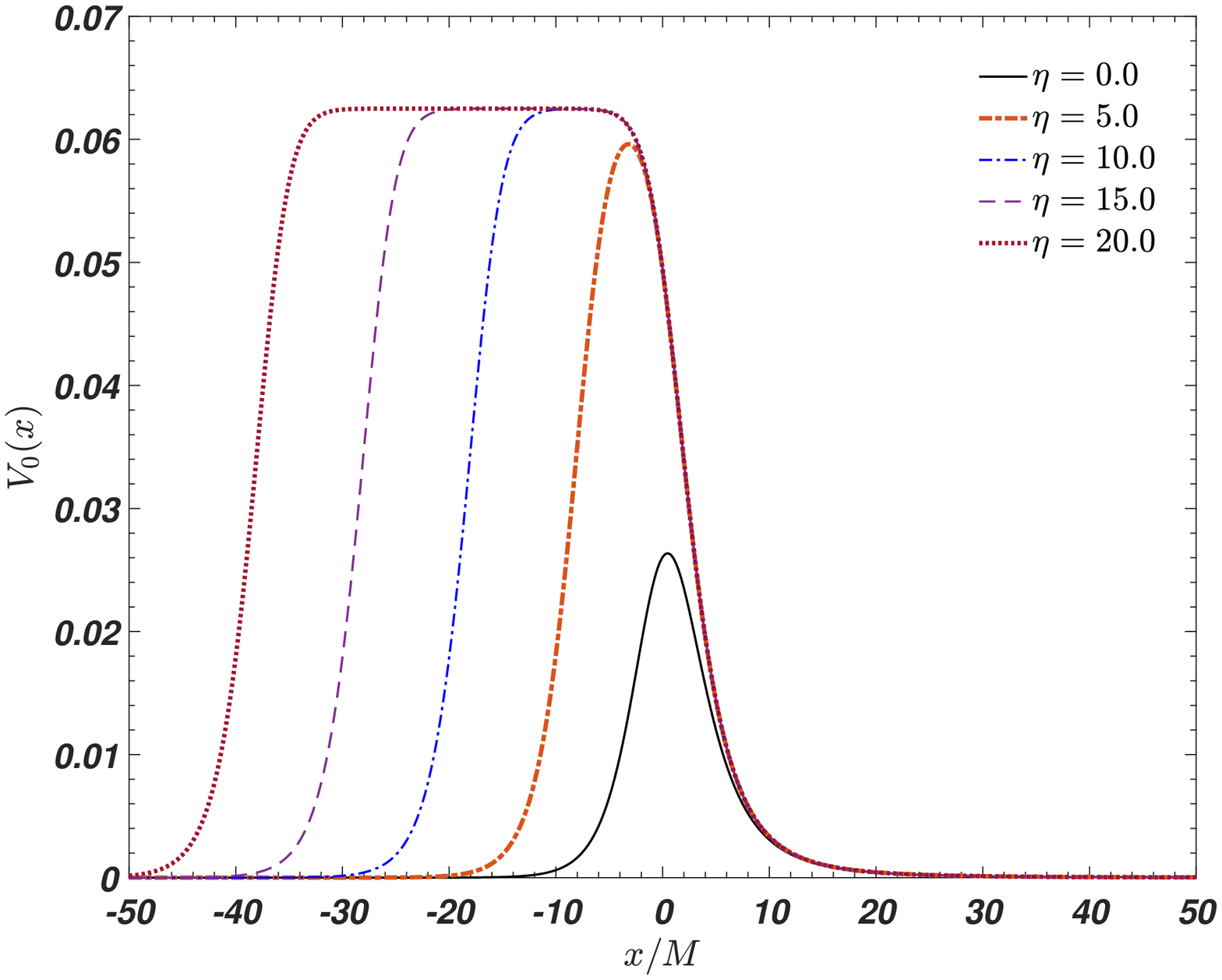}	
	\includegraphics[width=0.48\textwidth,height=0.38\textwidth]{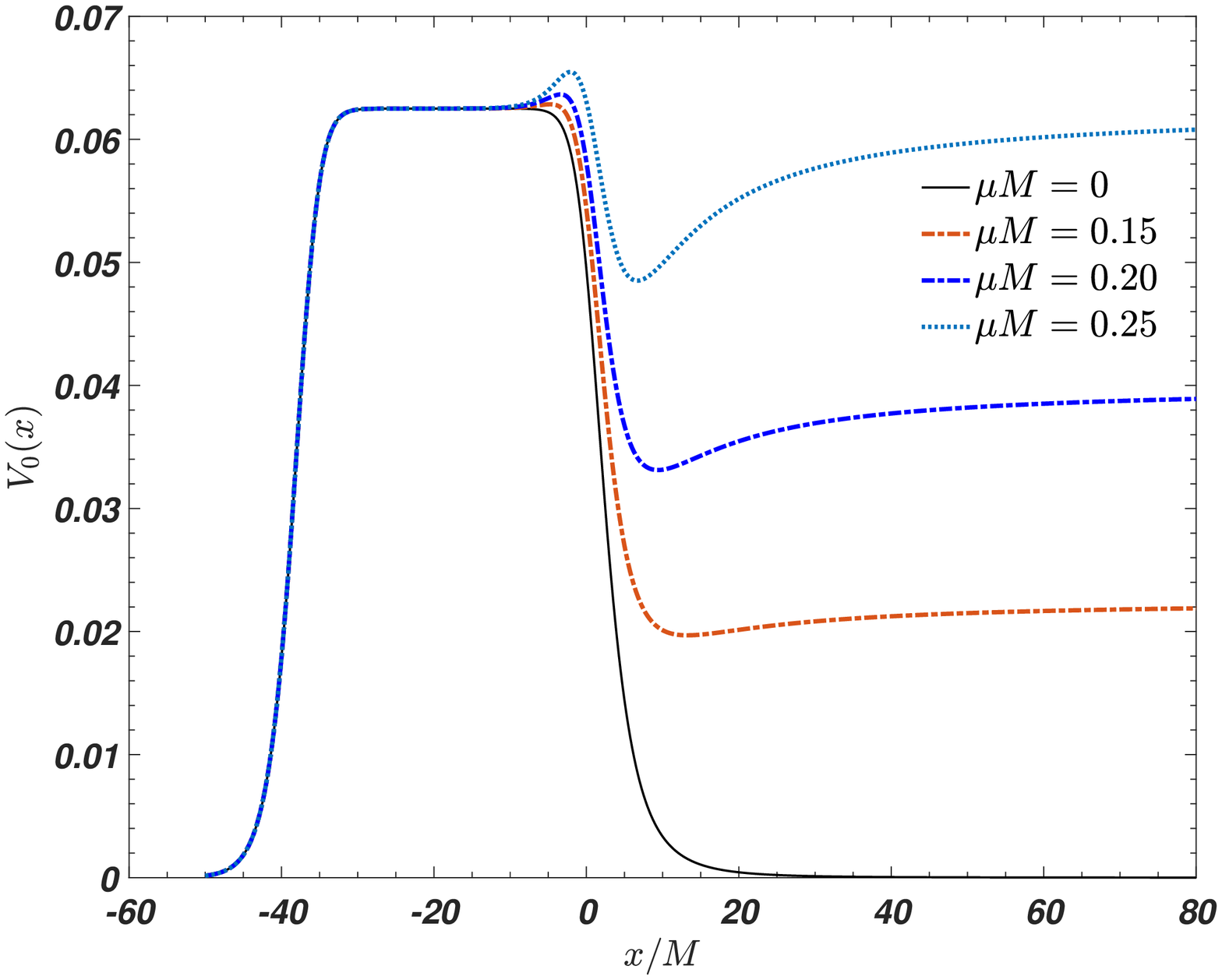}	
	\caption{Effective potential of $l=0$ state as functions of $x$. In the left panel, we fix $\mu M=0$ and compare the potential for different values of $\eta$, whereas in the right panel, we set $\eta=20$ and compare the potential for different values of $\mu M$.}
	\label{Fig: potential}
\end{figure*}

In Fig.\ref{Fig: potential}, we present the potential of $l=0$ state, as a function of $x$. The left panel shows the effect of $\eta$ on the potential. We fix $\mu M=0$ and gradually increase the values of $\eta$. In the Schwarzschild case ($\eta=0$), a potential barrier appears at $x/M\approx0$; For $\eta<5$ (or $q<0.993$), both the height and width of the potential increase with $\eta$; Interestingly, when $\eta>5$, the width continue to increase with $\eta$, while the height remains unchanged. In this case, the height of the potential is $(2l+1)^2/(16M^2)$.

\begin{figure}[!h]
	\centering	
	\includegraphics[width=0.45\textwidth,height=0.35\textwidth]{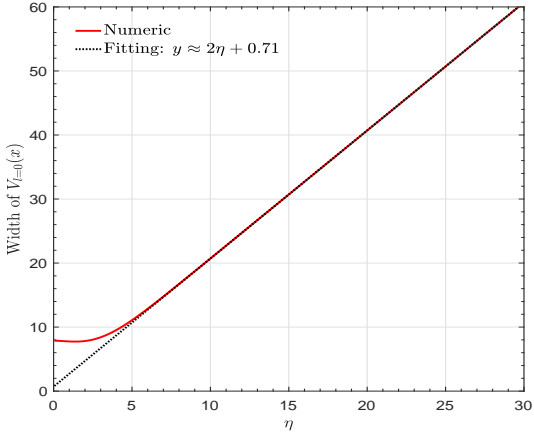}		
	\caption{Width of the effective potential with $l=0$ as a function of $\eta$.}
	\label{Fig: width of potential}
\end{figure}

To be more concrete, we define the width of the effective potential by $|x_2-x_1|$, where $x_{1,2}$ are the tortoise coordinate, at which $V(x=x_{1,2})=V_{\text{max}}/2$, where $V_{\text{max}}$ is the maximum value of the potential barrier.
In Fig. \ref{Fig: width of potential}, we present the width of the potential as a function of $\eta$. We see that the width is indeed increased monotonically with $\eta$. Especially, such growth is uniform when $\eta>5$.
Therefore, in the limit of $\eta\rightarrow\infty$, the GMGHS black hole becomes extremal and the potential barrier can be infinitely wide. This property is crucial for the existence of long-lived modes of massive scalar field adhered to a GMGHS black hole.

The right panel of Fig.\ref{Fig: potential} compares  potentials for different values of $\mu M$. For $\mu M\neq0$, the potential tends to $\mu^2$ at infinity, and there is a potential well between the barrier and spatial infinity. A wave with $\omega<\mu$ will bounce back and forth in the well, leaking its energy to the black hole each time due to the tunneling effect of the potential barrier. Such a tunneling effect will be heavily suppressed in the near extremal limit $\eta\gg1$, since the width of the potential barrier can be infinitely wide in this limit, as we have shown in the left panel in this figure. Thus, we expect that a surrounding massive scalar field could be notably long-lived for a near extremal GMGHS black hole.

\subsection{Quasibound states}
By imposing appropriate boundary conditions at the horizon and infinity, Eq.(\ref{Eq: the radial eq psi}) defines an eigenvalue problem of $\omega$.
Solutions of Eq.(\ref{Eq: the radial eq psi}) with boundary conditions (\ref{Eq: bcs infty}) and (\ref{Eq: bcs horizon}) are called quasibound states of a massive scalar field.
The two boundary conditions select a discrete set of complex frequencies (expressed by $\omega=\omega_R+i\omega_I$), which are called quasibound state frequencies.
The unstable mode corresponds to $\omega_I>0$, whereas $\omega_I<0$ corresponds to the stable mode with an e-folding decaying timescale: $\tau=|\omega_I|^{-1}$.

Previous studies have shown that in the limit $\mu M\ll l$, the real part of the the quasibound state frequency resembles that of the hydrogen atom
\begin{equation}\label{Eq: hydrogen}
	\omega_{n}\approx\left(1-\frac{\mu^2M^2}{2\tilde{n}^2}\right)\mu,
\end{equation}
where $\tilde{n}=n+l+1$ is the principal quantum number of the state. This formula also applies to the case of a massive scalar field around the GMGHS black hole.

In the following text, we shall derive a more accurate frequency spectrum of the quasibound states of a massive scalar field around the GMGHS black hole. More prominently, we shall show that the imaginary part of frequency tends to zero for a near extremal black hole ($\eta\rightarrow\infty$).

\section{Numerical method}
In this work, we apply the continued fraction method to compute the quasibound state frequencies.
In \cite{PhysRevD.92.064022}, the authors have used this method to find the quasinormal modes of massive scalar field in Kerr-Sen black hole spacetime. In this article, we apply it under quite different boundary conditions at infinity.

As discussed in Sec.\ref{Sec: setup}, for quasibound states the radial function decays exponentially at infinity.
Thus we may write the radial solution as
\begin{equation}
	\begin{aligned}
	R_{\omega l}(r)=\left(\frac{r-r_+}{r-r_-}\right)^{-i\sigma}&\left(r-r_-\right)^{\chi-1}e^{\rho r}\\&\times \sum_{n=0}^{\infty}a_{n}\left(\frac{r-r_+}{r-r_-}\right)^{n},
	\end{aligned}
\end{equation}
where $\rho=-\sqrt{\mu^2-\omega^2}$, $\chi=M(\mu^2-2\omega^2)/\rho$, and $\sigma=2M\omega$. Substituting this equation into Eq.(\ref{Eq: the radial eq}) yields a three-term recurrence relation among the expansion coefficients $a_n$
\begin{subequations}
	\begin{empheq}{align}
	\alpha_0a_1+\beta_0a_0 &=0,\\
	\alpha_na_{n+1}+\beta_na_n+\gamma_na_{n-1} &=0,\;\;\;\text{for}\;\;\;n>0,
	\end{empheq}
\end{subequations}
where $\alpha_n$, $\beta_n$ and $\gamma_n$ are respectively given by
\begin{equation}
	\alpha_n=\left(1+n\right)\left(1+n-4i\omega\right),
\end{equation}
\begin{equation}
	\begin{aligned}
	\beta_n=&-2n^2-2\left[1-4i\omega+(2q^2-3)\rho+\frac{\omega^2}{\rho}\right]n\\
		    &-l(l+1)-\left(1-4i\omega\right)\left[1+(2q^2-3)\rho+\frac{\omega^2}{\rho}\right]\\
		    &+8\omega^2+4(1-q^2)(2\omega^2-\mu^2),
	\end{aligned}
\end{equation}
and
\begin{equation}
	\gamma_n=n^2-4i\omega n-8\omega^2+\frac{2(n-2i\omega)(2\omega^2-\mu^2)}{\rho}+\frac{\mu^4}{\rho^2}.
\end{equation}
Here, we set $M=1$, and all other quantities are measured by $M$. Then, the quasibound state frequencies are obtained by solving numerically the following equation \cite{Leaver285}
\begin{equation}
	0=\beta_0-\frac{\alpha_0\gamma_1}{\beta_1-}\frac{\alpha_1\gamma_2}{\beta_2-}\frac{\alpha_2\gamma_3}{\beta_3-}\cdots.
\end{equation}

\section{Analytical method}\label{Sec: analytic}
In the near-extremal limit ($q\rightarrow1$), the eigenvalue problem introduced in previous section can be solved analytically. It is convenient to introduce the following dimensionless quantities
\begin{equation}
z\equiv\frac{r-r_+}{r_+};\;\;\;\tau\equiv\frac{r_+-r_-}{r_+}=1-q^2.
\end{equation}
Then, the radial equation (\ref{Eq: the radial eq}) becomes
\begin{equation}\label{Eq: the radial eq dimensionless}
z(z+\tau)\frac{d^2R}{dz^2}+(2z+\tau)\frac{dR}{dz}+VR=0,
\end{equation}
with
\begin{equation}
	\begin{aligned}
	V=&\frac{4\tau\epsilon^2}{z}-l(l+1)+4(1+2\tau)\epsilon^2-4\tau\mu_s^2\\&+4\left[\epsilon^2+(1+\tau)(\epsilon^2-\mu^2_s)\right]z+4(\epsilon^2-\mu^2_s)z^2.
	\end{aligned}
\end{equation}
where $\epsilon=\omega M$ and $\mu_s=\mu M$ are the dimensionless frequency and mass respectively.

\subsection{Near horizon solution}
Close to the event horizon $z\rightarrow0$, we can omit higher order terms of $z$. Hence, Eq.(\ref{Eq: the radial eq dimensionless}) becomes
\begin{equation}\label{Eq: the radial eq dimensionless near}
z(z+\tau)\frac{d^2R}{dz^2}+(2z+\tau)\frac{dR}{dz}+\left(\frac{4\tau\epsilon^2}{z}+\frac{1}{4}-\beta^2\right)R=0,
\end{equation}
where
\begin{equation}\label{Eq: beta}
\beta^2=\left(l+\frac{1}{2}\right)^2+4\tau\mu_s^2-4(1+2\tau)\epsilon^2.
\end{equation}
The ingoing wave solution of Eq.(\ref{Eq: the radial eq dimensionless near}) is
\begin{equation}
R(z)\sim\left(\frac{z}{\tau}\right)^{-2i\epsilon}\;_2F_1\left(\frac{1}{2}-\beta-2i\epsilon,\frac{1}{2}+\beta-2i\epsilon;1-4i\epsilon;-\frac{z}{\tau}\right),
\end{equation}
where $_2F_1(a,b;c;z)$ is the hypergeometric function. Considering the limit $z\gg\tau$, and using the property of the hypergeometric function
\begin{equation}
_2F_1(a,b;c;z)=\frac{\Gamma(c)\Gamma(b-a)}{\Gamma(c-a)\Gamma(b)}u_3+\frac{\Gamma(c)\Gamma(a-b)}{\Gamma(c-b)\Gamma(a)}u_4,
\end{equation}
where
\begin{subequations}
	\begin{empheq}{align}
	u_{3}&=(-z)^{-a}\;_2F_1\left(a,a+1-c;a+1-b;\frac{1}{z}\right),\\
	u_{4}&=(-z)^{-b}\;_2F_1\left(b,b+1-c;b+1-a;\frac{1}{z}\right),
	\end{empheq}
\end{subequations}
the near horizon solution can be written as
\begin{equation}\label{Eq: Near sol for large z}
R(z)\sim \frac{\Gamma(1-4i\epsilon)\Gamma(2\beta)}{\Gamma\left(\frac{1}{2}+\beta-2i\epsilon\right)^2}\left(\frac{z}{\tau}\right)^{-\frac{1}{2}+\beta}+(\beta\rightarrow-\beta).
\end{equation}
Here the second term $(\beta\rightarrow-\beta)$ refers to a replacement of $\beta$ with $-\beta$ of the first term.

\subsection{Far region solution}
Now we consider the far region solution of the radial equation. For $z\gg\tau$, Eq.(\ref{Eq: the radial eq dimensionless}) becomes
\begin{equation}\label{Eq: the radial eq dimensionless far}
z^2\frac{d^2R}{dz^2}+2z\frac{dR}{dz}+\left[\frac{1}{4}-\beta^2+2\kappa kz-k^2z^2\right]R=0,
\end{equation}
where $\beta$ is given in Eq.(\ref{Eq: beta}), and
\begin{equation}
\kappa=\frac{4\epsilon^2-(1+\tau)k^2}{2k},
\end{equation}
with $k=2\sqrt{\mu^2_s-\epsilon^2}$. The solution of Eq.(\ref{Eq: the radial eq dimensionless far}) is
\begin{equation}\label{Eq: sol far}
\begin{aligned}
R(z)=&C_1e^{-kz}(2k)^{\frac{1}{2}+\beta} z^{-\frac{1}{2}+\beta}M\left(\frac{1}{2}+\beta-\kappa,1+2\beta,2kz\right)\\&+C_2\times\left(\beta\rightarrow-\beta\right),
\end{aligned}
\end{equation}
where $M(a,c,z)$ is the confluent hypergeometric function, and $\left\lbrace C_1,C_2\right\rbrace$ are constants to be determined by the matching and boundary conditions.
For $z\gg1$, the confluent hypergeometric function behaves as
\begin{equation}
M(a,c,z)\sim\frac{\Gamma(c)}{\Gamma(a)}e^{z}z^{a-c}+\frac{\Gamma(c)}{\Gamma(c-a)}(-1)^az^{-a}.
\end{equation}
Hence, for $z\gg1$, the far region solution (\ref{Eq: sol far}) reduces to
\begin{equation}\label{Eq: sol at infty}
\begin{aligned}
R(z\rightarrow&\infty)\sim\\
&\Bigg[
C_1\frac{\Gamma(1+2\beta)}{\Gamma\left(\frac{1}{2}+\beta-\kappa\right)}+C_2\times(\beta\rightarrow-\beta)
\Bigg]\times\\&(2k)^{-\kappa}z^{-1-\kappa}e^{kz}+\\&
\Bigg[
C_1\frac{\Gamma(1+2\beta)}{\Gamma\left(\frac{1}{2}+\beta+\kappa\right)}(2k)^{\kappa}z^{-1+\kappa}(-1)^{\frac{1}{2}+\beta-\kappa}\\&
+C_2\times(\beta\rightarrow-\beta)\Bigg]
e^{-kz}.
\end{aligned}
\end{equation}
For quasibound states, the radial function tends to zero at infinity. Therefore, the coefficient of the first term in Eq.(\ref{Eq: sol at infty}) equals to zero
\begin{equation}\label{Eq: boundstate condition 1}
C_1\frac{\Gamma(1+2\beta)}{\Gamma\left(\frac{1}{2}+\beta-\kappa\right)}
+C_2\times\left(\beta\rightarrow-\beta\right)=0.
\end{equation}

\subsection{Matching the two solution}
For near extremal GMGHS black holes with $\tau\ll1$, there is an overlap region $\tau\ll z\ll1$ in which the two solutions should match each other.
We have obtained the $z\gg\tau$ behavior of the near horizon solution, see Eq.(\ref{Eq: Near sol for large z}). On the other hand, the $z\ll1$ limit
of the far region solution (\ref{Eq: sol far}) is
\begin{equation}
R(z)\sim C_1(2k)^{\frac{1}{2}+\beta}z^{-\frac{1}{2}+\beta}+C_2\times\left(\beta\rightarrow-\beta\right).
\end{equation}
Comparing this equation with Eq.(\ref{Eq: Near sol for large z}), we obtain
\begin{equation}
\begin{aligned}
&C_1(\beta)=\left(\frac{\tau}{2k}\right)^{1/2}\frac{\Gamma(1-4i\epsilon)\Gamma(+2\beta)}{\Gamma\left(\frac{1}{2}+\beta-2i\epsilon\right)^2}(2k\tau)^{-\beta},\;\text{and}\\&
C_2(\beta)=C_1(-\beta).
\end{aligned}
\end{equation}
Substituting this equation into Eq.(\ref{Eq: boundstate condition 1}), we have
\begin{equation}\label{Eq: boundstate condition 2}
	\frac{\Gamma\left(\frac{1}{2}-\beta-\kappa\right)}{\Gamma\left(\frac{1}{2}+\beta-\kappa\right)}=\frac{\Gamma\left(\frac{1}{2}+\beta-2i\epsilon\right)^2\Gamma\left(-2\beta\right)^2}{\Gamma\left(\frac{1}{2}-\beta-2i\epsilon\right)^2\Gamma\left(+2\beta\right)^2}\left(2k\tau\right)^{2\beta}.
\end{equation}
This is the equation of the quasibound state frequency in the near extremal limit. For given values of $\left\lbrace l,\mu,\tau\right\rbrace $, we compute the eigenvalues of $\epsilon$ by solving Eq.(\ref{Eq: boundstate condition 2}) numerically. If $\epsilon<\mu_s<L/2$, this equation can also be solved iteratively.
For $\tau\ll1$, the right hand side of Eq.(\ref{Eq: boundstate condition 2}) is very small. Therefore, we obtain the approximation of $\epsilon$ by requiring that the left hand side equals to zero. Using the property of the Gamma function $1/\Gamma(-n)=0$, we have
\begin{equation}\label{Eq: boundstate condition 3}
\frac{1}{2}+\beta-\kappa=-n.
\end{equation}
This equation can be solved by assuming
\begin{equation}\label{Eq: SORP}
\epsilon=\mu_s\left[1+\sum_{i=1}^{\infty}C_i\mu_s^{2i}\right].
\end{equation}
Substituting Eq.(\ref{Eq: SORP}) into Eq.(\ref{Eq: boundstate condition 3}), the coefficients $C_i$ can be solve order by order for arbitrary $i$. Here we list the first third $C_i$:
\begin{subequations}\label{Eq: Ci}
	\begin{empheq}{align}
	C_1 &=-\frac{1}{2\tilde{n}^2},\\
	C_2 &=-\frac{2(1+\tau)}{\tilde{n}^3L}+\frac{\tau +15/8}{\tilde{n}^4},\\
	C_3 &=-\frac{2(\tau+1)^2}{\tilde{n}^3L^3}-\frac{6(\tau +1)^2}{\tilde{n}^4L^2}\\&+\frac{8\tau^2+27\tau +17}{\tilde{n}^5L}-\frac{40 \tau ^2+152 \tau +145}{16\tilde{n}^6},
	\end{empheq}
\end{subequations}
where $\tilde{n}=n+l+1$ and $L=l+1/2$. For long-lived modes with $\omega_I\ll\omega_R$, Eq.(\ref{Eq: SORP}) may be treated as an approximation of the real part $\omega_R$. Clearly, in the small mass limit $\mu_s\ll1$, we can omit $C_i$ for $i\geq2$, and recover the hydrogenic spectrum, see Eq.(\ref{Eq: hydrogen}). However, if $\mu_s$ is comparable to $L$, i.e, $\mu_s\lesssim L/2$, it is necessary to consider higher order terms to get more accurate results.

Comparing to the real part, we derive the imaginary part via a different way. The imaginary part $\omega_I$ can be obtained perturbatively. Substituting $\epsilon=\omega_R+i\omega_I$ into Eq.(\ref{Eq: boundstate condition 2}), we obtain the equation for $\omega_I$. Then, we treat both $\omega_I$ and the right hand side of Eq.(\ref{Eq: boundstate condition 2}) as small numbers, and expand the equation in terms of $\omega_I$. Finally, we obtain the approximation of $\omega_I$ by letting the linear part of the expansion equals to zero. The whole procedure is tediously lengthy, and the resulting formula is too cumbersome to be presented here. In brief, we find
\begin{equation}
	\omega_IM\propto-e^{-2\tilde{\beta}\eta},
\end{equation}
where $\eta$ is defined in Eq.(\ref{Eq: eta}) and
\begin{equation}
	\tilde{\beta}=\sqrt{\left(l+\frac{1}{2}\right)^2-4\omega_R^2}.
\end{equation}

\section{Results}
In this section, we systematically demonstrate our numerical and analytical results of the frequencies of quasibound states, both for ground states and excited states.
Significantly, we find that the numerical results agree well with the analytical results, especially for near extremal black holes.

\begin{figure*}
	\centering	
	\includegraphics[width=0.45\textwidth,height=0.35\textwidth]{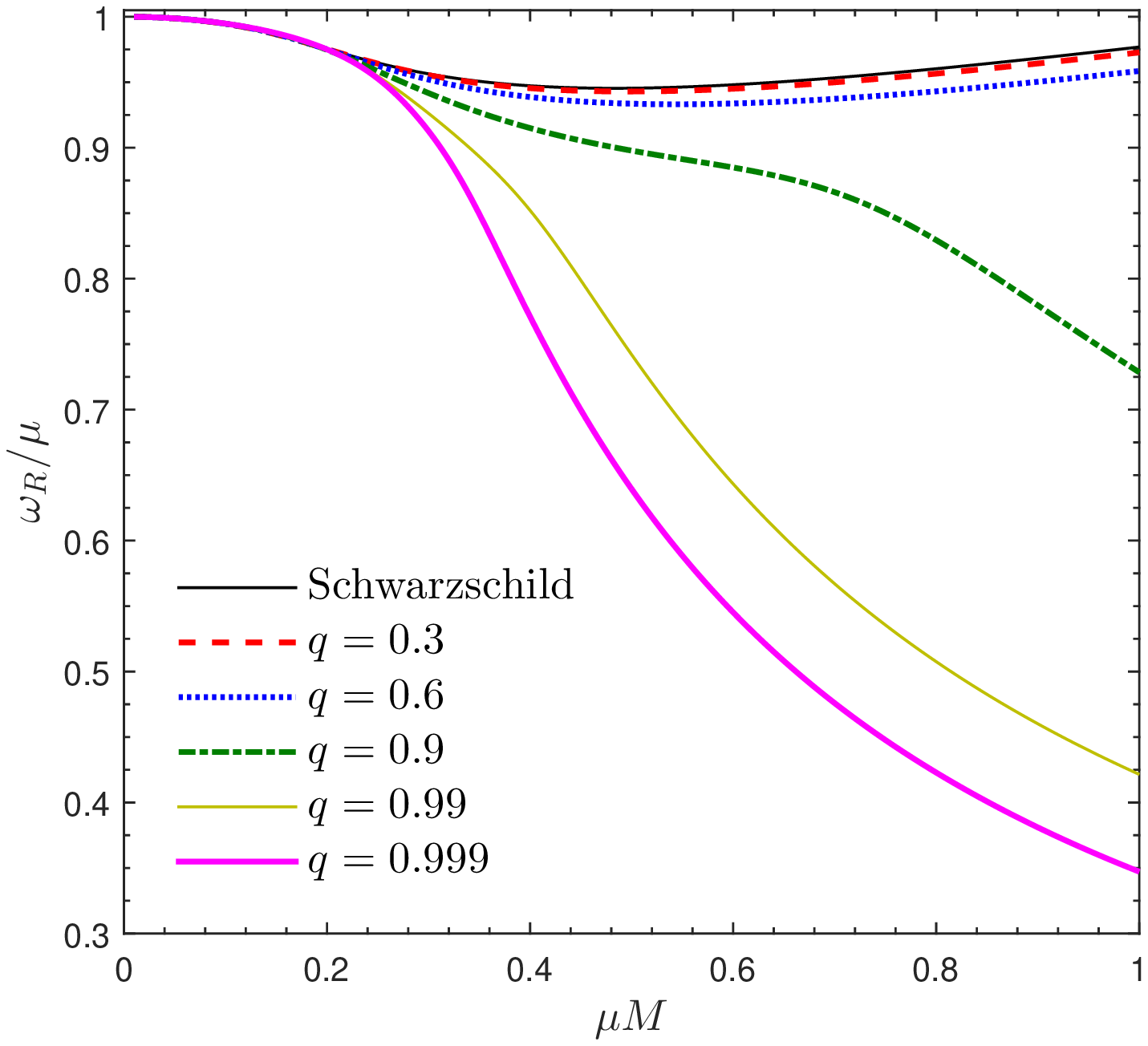}
	\includegraphics[width=0.45\textwidth,height=0.35\textwidth]{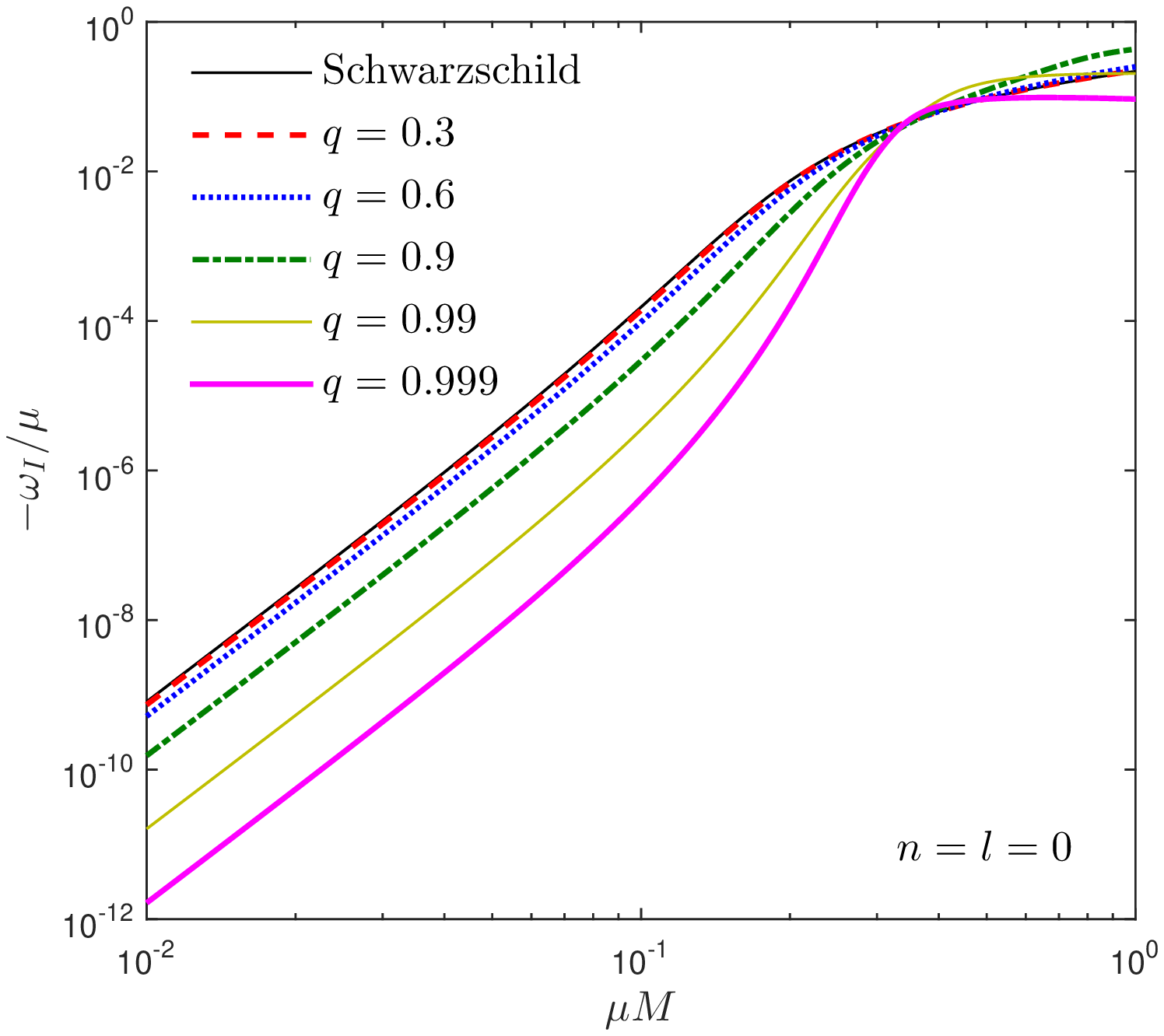}			
	\caption{The  frequencies of ground quasibound state of $l=0$ state as functions of $\mu M$. The left panel shows the real part $\omega_R$, whereas the right panel shows the corresponding lines for the imaginary part $\omega_I$.}
	\label{Fig: quasibound states}
\end{figure*}

First, we show the complex frequencies of  ground quasibound state of GMGHS black hole with different charges varies with respect to fine structure constants $\mu M$. Figure \ref{Fig: quasibound states} compares the spectra of $l=0$ states for different values of black hole charge $q$. Here, the Schwarzschild case ($q=0$) is also plotted in comparison.
For a very small fine structure constant $\mu M\ll1$, the Compton wave length of the scalar field is much larger than the size of the black hole. In this case, the real part of the frequency is well described by Eq.(\ref{Eq: hydrogen}), and the effect of $q$ is negligible. Physically, a very long wave hardly senses the existence of charge. Similar phenomena appear in scattering of GMGHS black holes and Reissner-Nordstrom black holes \cite{Huang:2020bdf}. As the mass coupling $\mu M$ is increased, the effect of $q$ on $\omega_R$ becomes significant. The physical explanation is that a shorter wave length yields a smaller expectation of spatial radius of the wave function, which implies that the surrounding massive particle is nearer to the central black hole than the weak coupling case. The gravitational effect of the charge decreases with respect to $r$ much faster than that of the mass. Thus for a neutral particle far away from a charged mass point, it almost does not senses the gravity of charge. But when it is put nearer and nearer to the charged mass point, the gravity effects of the charge of the hole becomes more and more evident.

As for the imaginary part of the frequency, the effect of $q$ is significant, even in the limit $\mu M\ll1$. From the right panel of Fig.\ref{Fig: quasibound states}, we see that for given value of $\mu M$, $|\omega_I|$ decreases with the increase of $q$. Such trends is more notable in the near extremal limit. For example, when $\mu M=0.2$, the value of $|\omega_I|$ for $q=0.9$ is about $4$ times of that for $q=0.99$; And the value of $|\omega_I|$ for $q=0.99$ is again about $4$ times of that for $q=0.999$, whereas the increase of $q$ is only one tenth of the former case. Physically, from Fig. \ref{Fig: potential} it is clear that a larger $q$ leads to a wider potential. A wider potential is more difficult to penetrate, which is equal to say a smaller imaginary part of the frequency.

\begin{figure*}
	\centering	
	\includegraphics[width=0.45\textwidth,height=0.35\textwidth]{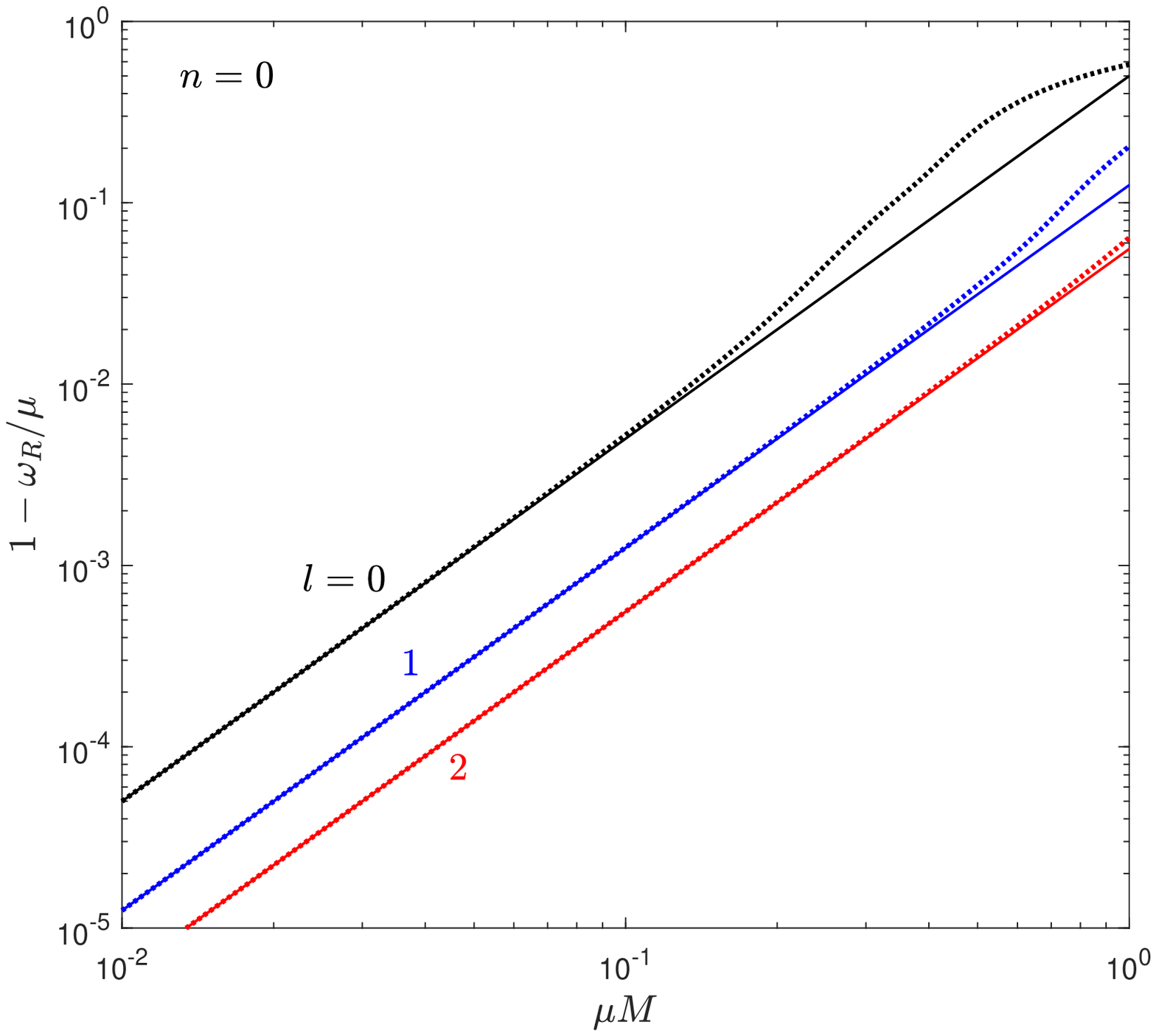}
	\includegraphics[width=0.45\textwidth,height=0.35\textwidth]{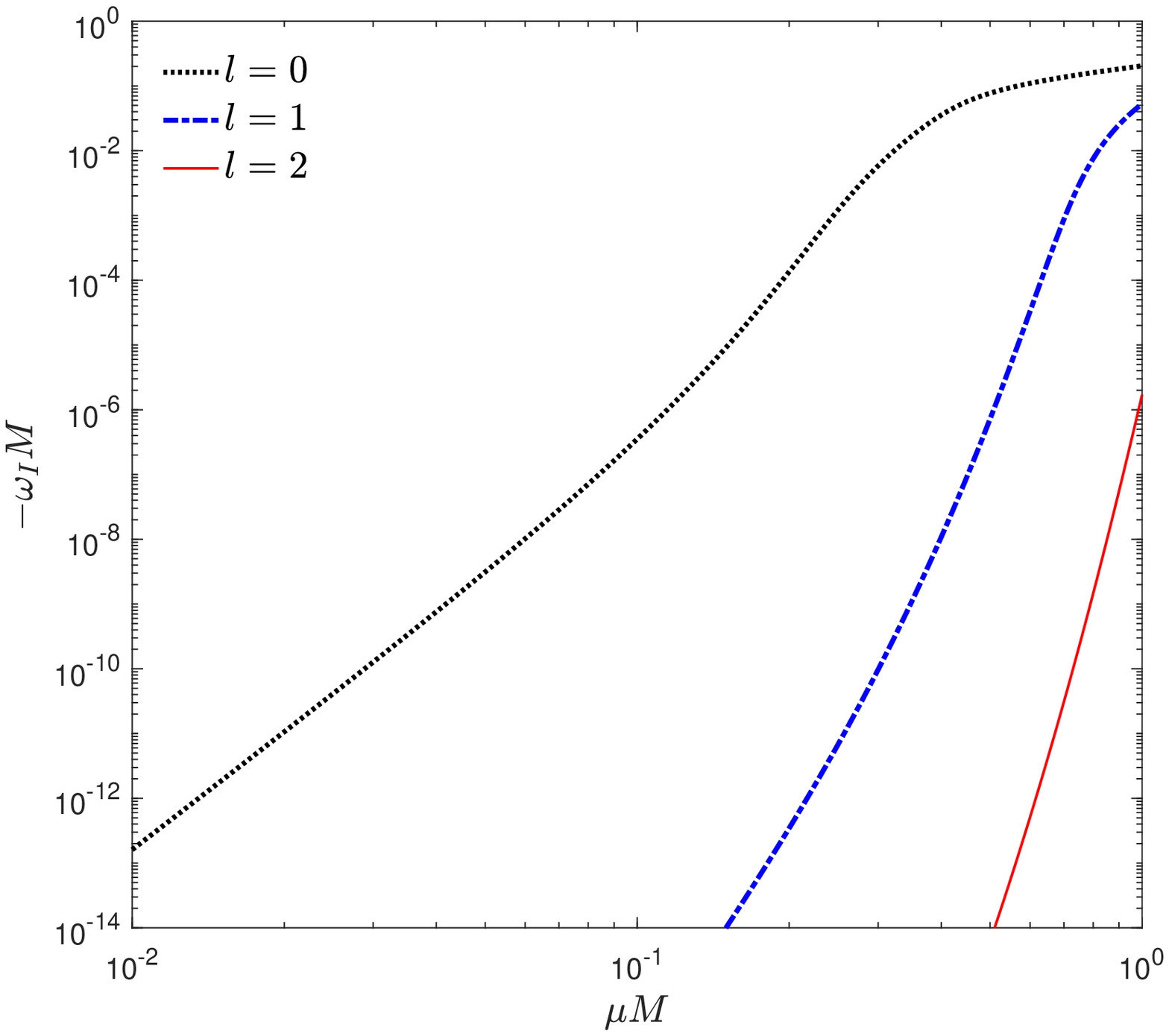}			
	\caption{Spectra of the quasibound states, with $n=0$, for $l=0,1$ and $2$. The black hole charge is $q=0.99$. The left panel shows the real part $\omega_R$ (or more precisely $1-\omega_R/\mu$), whereas the right panel shows the corresponding lines for the imaginary part $\omega_I$. In the left panel, the dashed lines denote numerical results, whereas the solid lines denote the analytic results, see Eq.(\ref{Eq: hydrogen}).}
	\label{Fig: vary l}
\end{figure*}

In Fig.\ref{Fig: vary l}, we present energy spectra for different angular momentum $l$. This figure shows that for small values of $\mu M$, the real part of the frequency is well approximated by Eq.(\ref{Eq: hydrogen}), and the imaginary part $\omega_I$ tends to zero in the limit $\mu M\ll1$. This means that ultralight scalar field can be very long-lived around the GMGHS black hole. Similar results have been found for massive scalar and Dirac fields around a Schwarzschild or a Reissner-Nordstr\"{o}m black hole\cite{Barranco:2012qs,Huang:2017nho,Zhou:2013dra}.

\begin{figure*}
	\centering	
	\includegraphics[width=0.32\textwidth,height=0.26\textwidth]{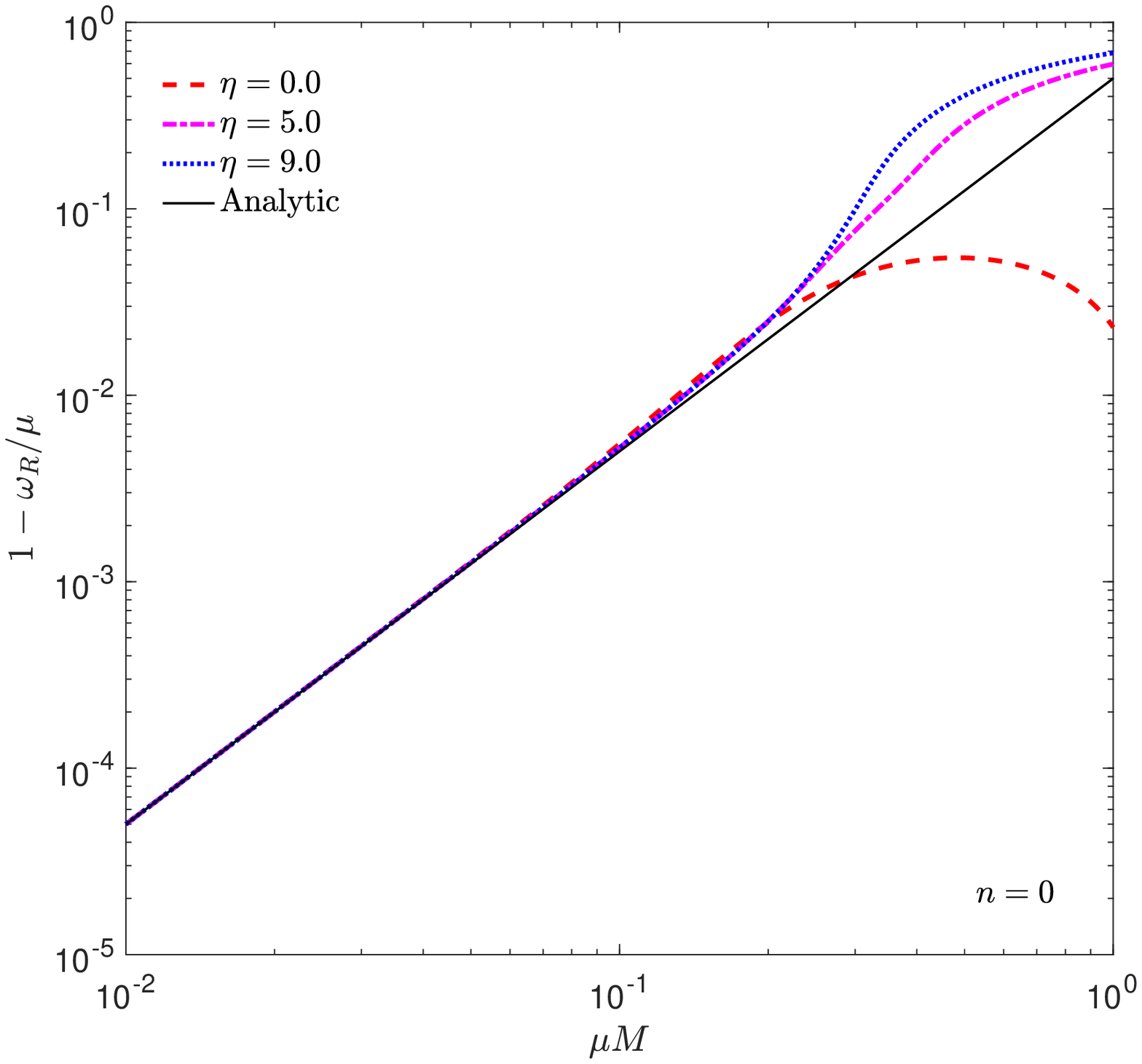}
	\includegraphics[width=0.32\textwidth,height=0.26\textwidth]{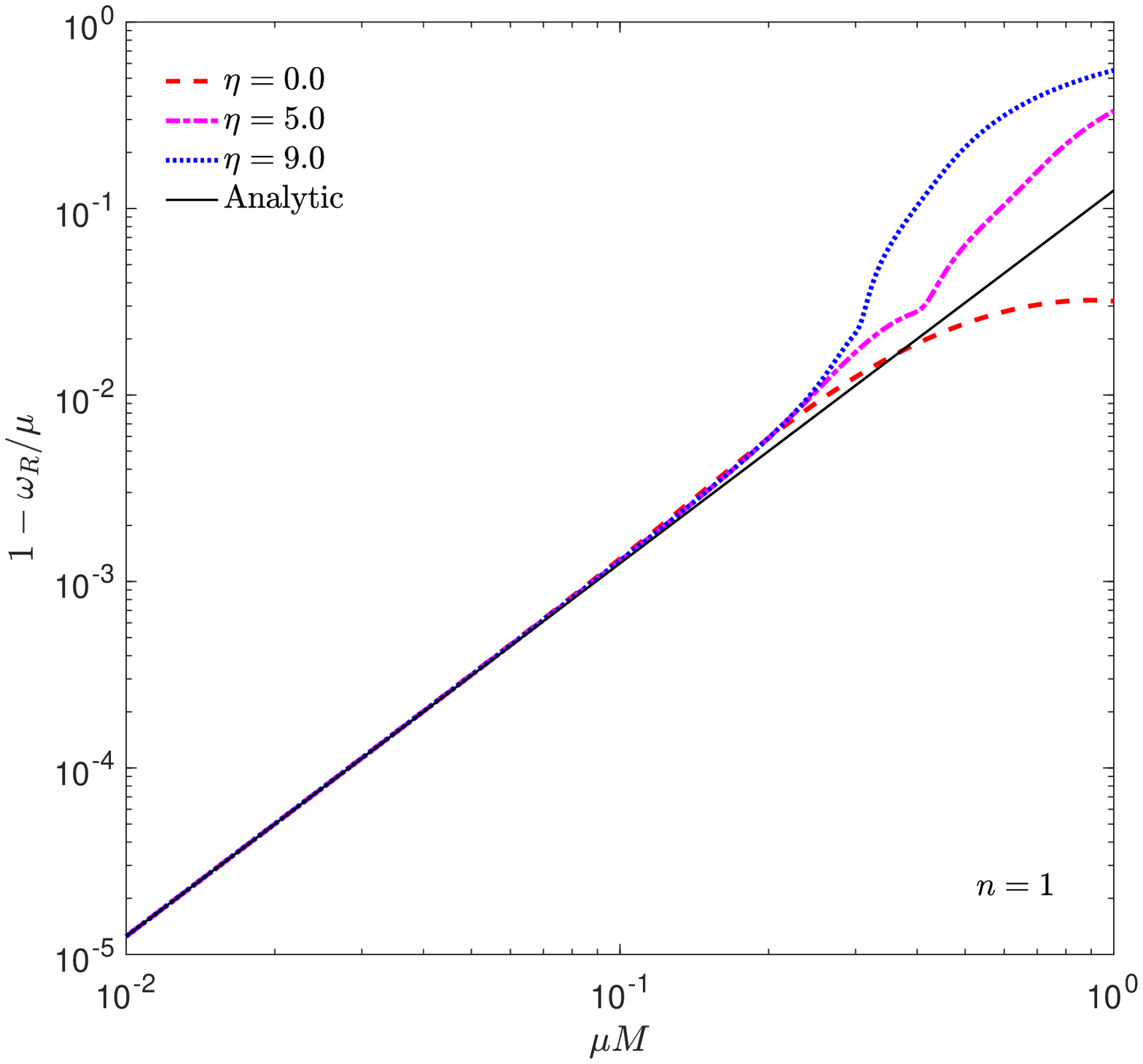}
	\includegraphics[width=0.32\textwidth,height=0.26\textwidth]{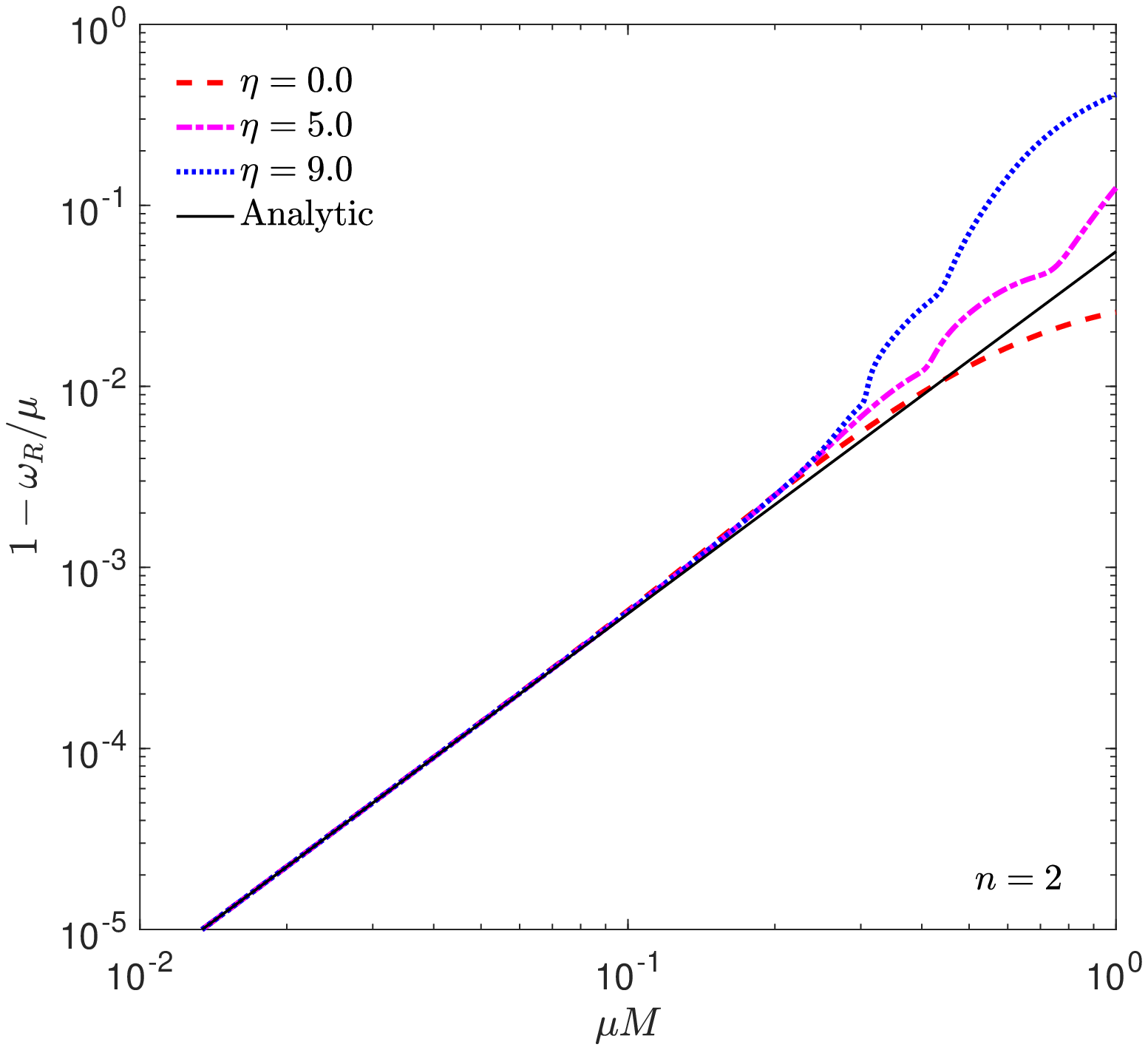}
	
	\includegraphics[width=0.32\textwidth,height=0.26\textwidth]{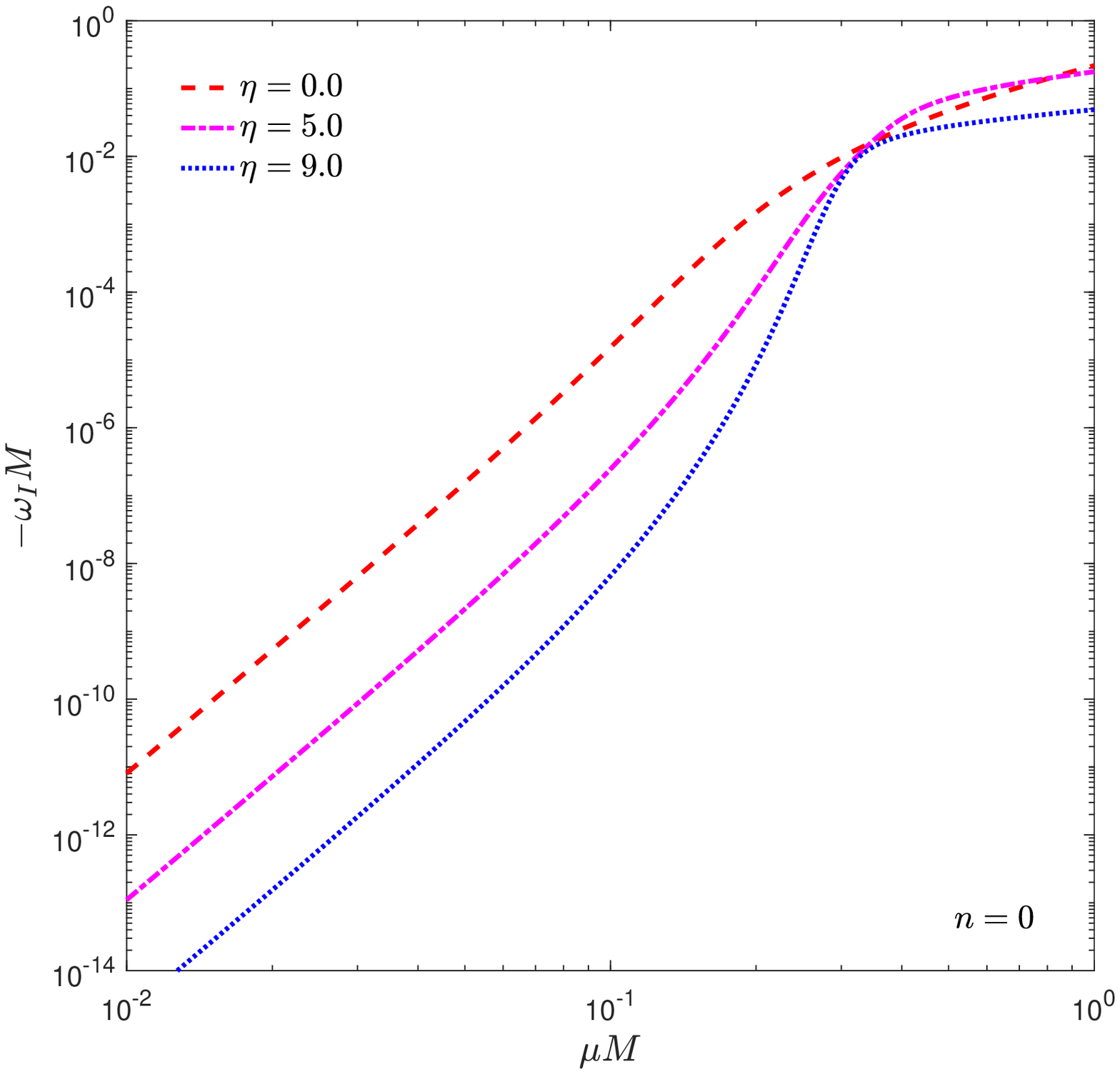}
	\includegraphics[width=0.32\textwidth,height=0.26\textwidth]{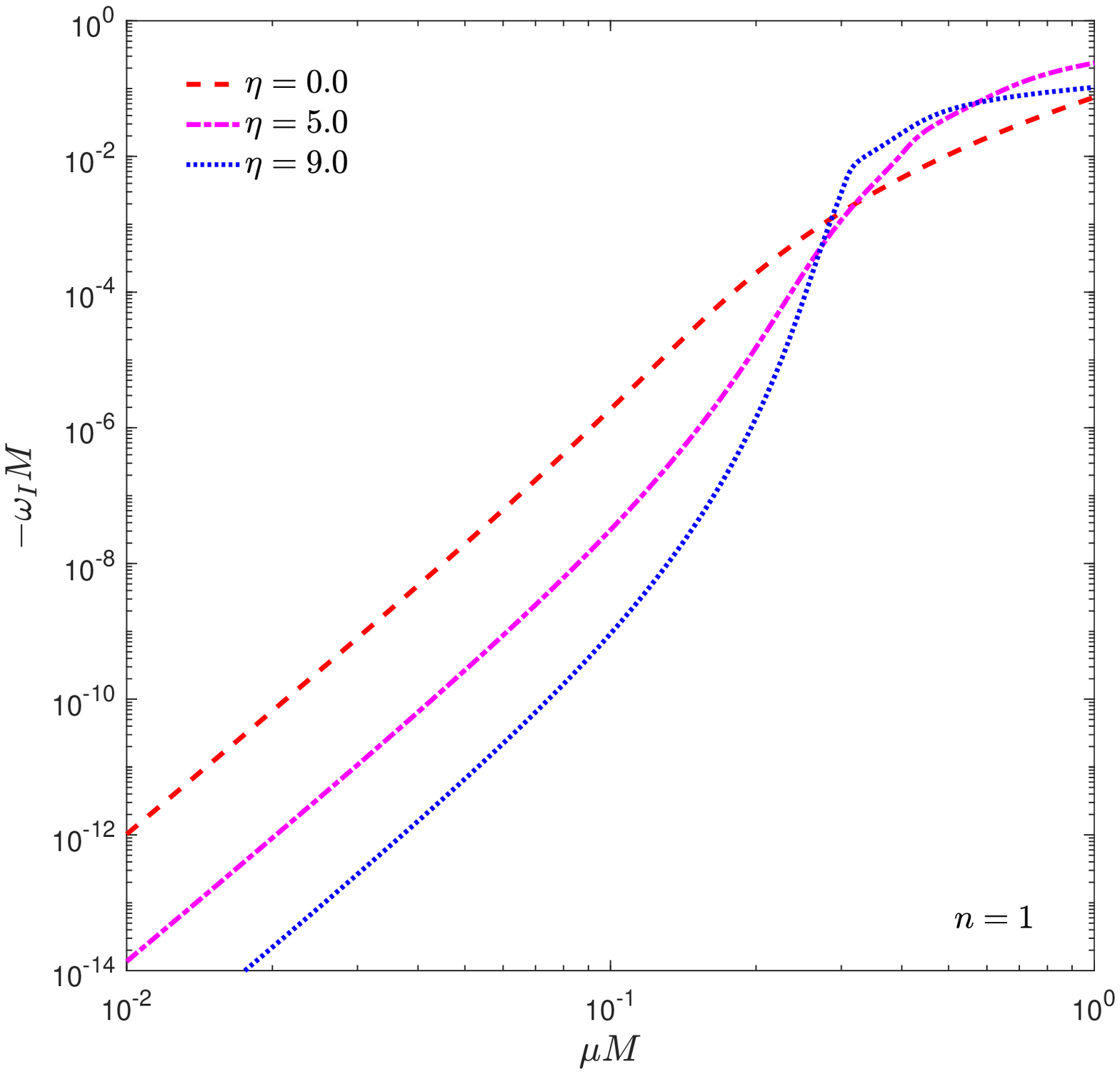}
	\includegraphics[width=0.32\textwidth,height=0.26\textwidth]{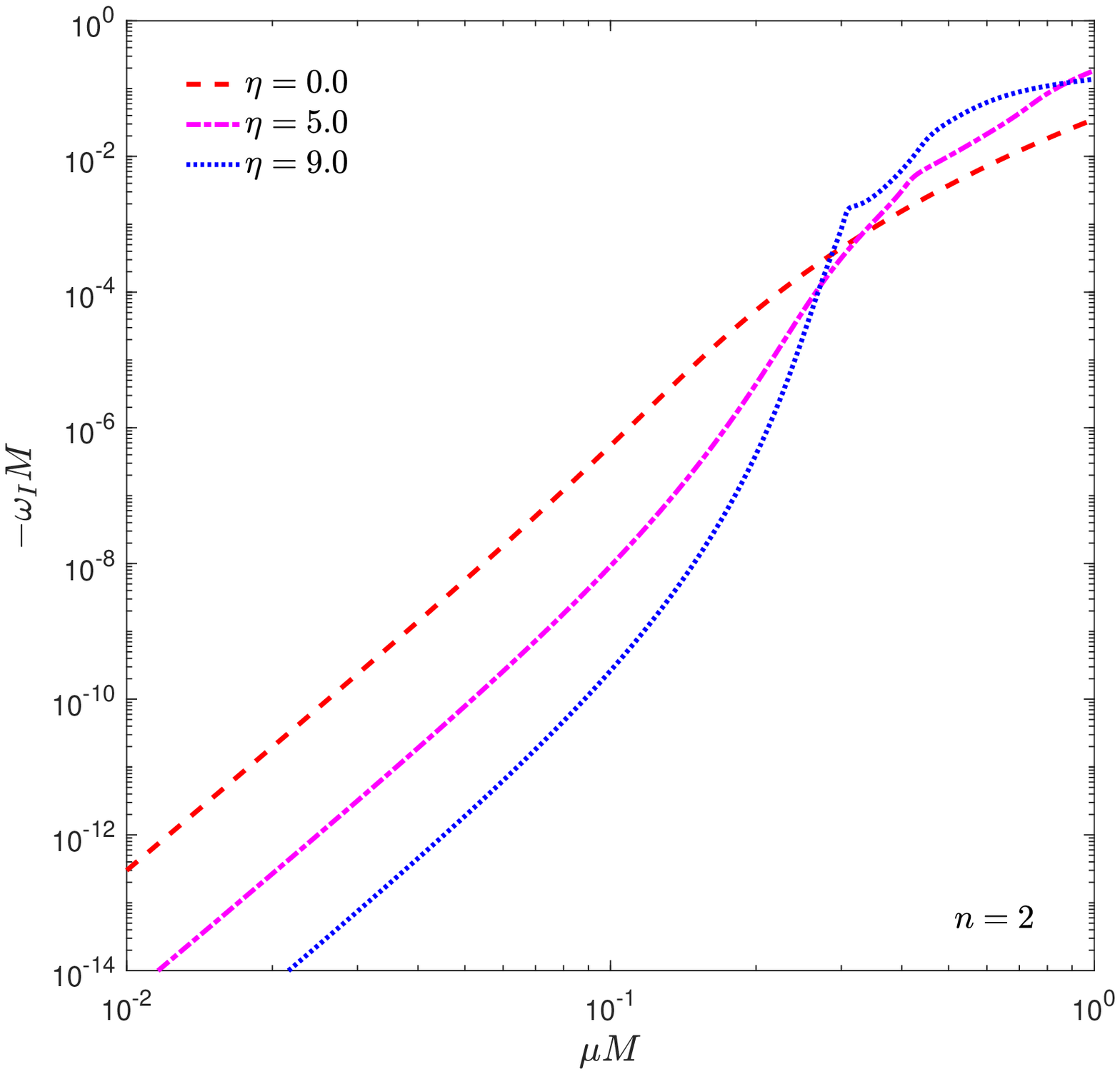}			
	\caption{Spectra of $l=0$ state for different values of $\eta$. The upper panel shows the real part $\omega_R$ (or more precisely $1-\omega_R/\mu$), whereas the bottom panel shows the corresponding lines for the imaginary part $\omega_I$. The analytical result in this figure is given by Eq.(\ref{Eq: hydrogen}). The excitation number $n=0,1$ and $2$ for the left, middle and right panels, respectively.}
	\label{Fig: quasibound states 2}
\end{figure*}

From the analysis in Sec.\ref{subSec: KG eq}, we expect that $\omega_I$ tends to zero in the extremal limit. Figure \ref{Fig: quasibound states 2} compares the spectra of $l=0$ state for different values of $\eta$.
The upper panel shows more clearly that the real part of the frequency of the quasibound states with the same value of $\tilde{n}=n+l+1$ degenerate in the limit $\mu M\ll1$ and such degeneracy does not depend on the black hole charge, as predicted by Eq.(\ref{Eq: hydrogen}). From Eqs.(\ref{Eq: SORP}) and (\ref{Eq: Ci}), the parameter $\eta$ only affects $\omega_R$ when the coupling constant $\mu M$ is comparable to $L$. The effects of $\eta$ on $\omega_I$ is much more notable. We see that for a given $\mu M$, $|\omega_I|$ decreases with the increase of $\eta$. This confirms our physical analysis from Fig.\ref{Fig: quasibound states}.

\begin{figure}
	\centering	
	\includegraphics[width=0.45\textwidth,height=0.36\textwidth]{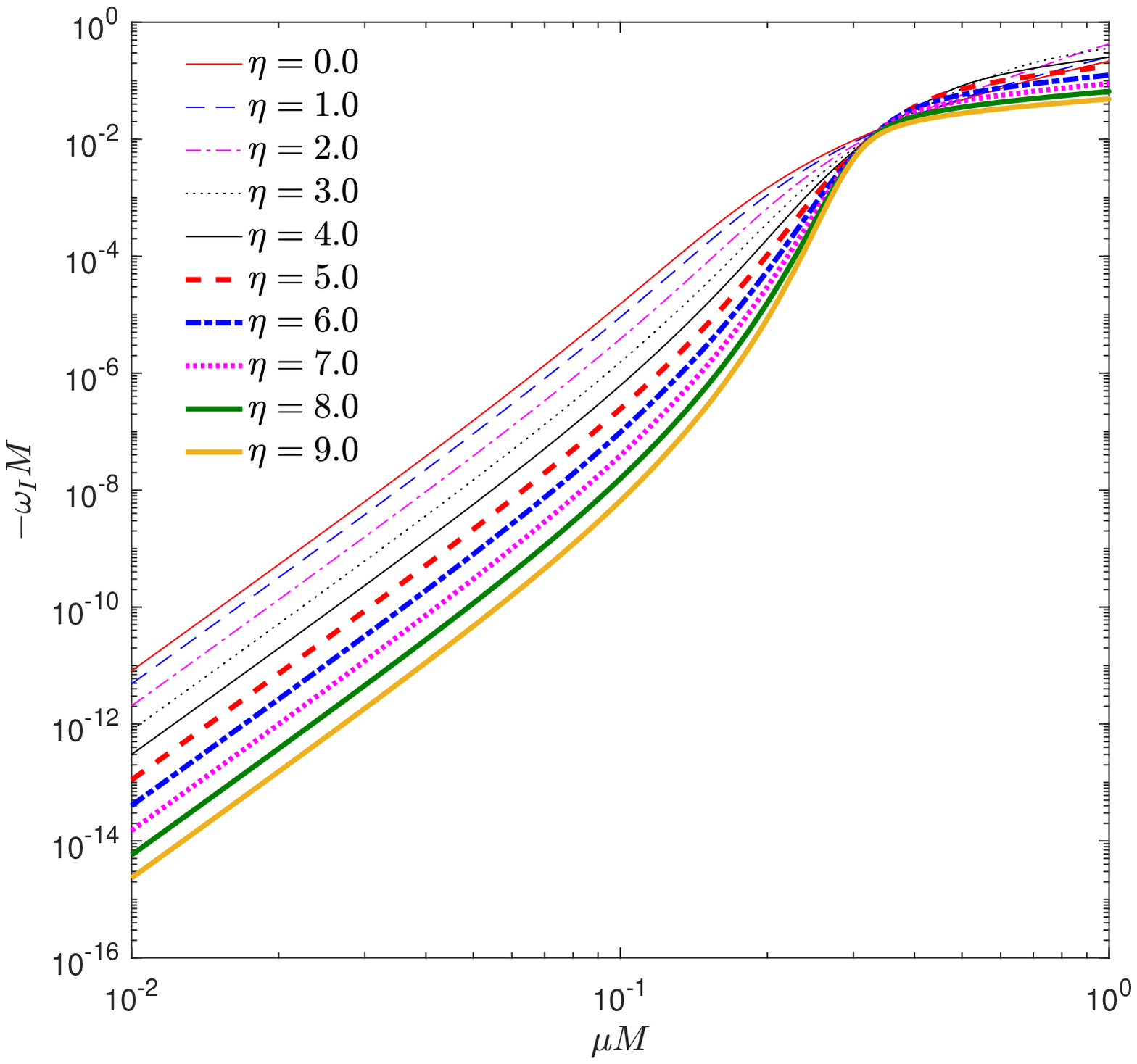}		
	\caption{Imaginary part of the quasibound state frequencies, with $n=l=0$, as a function of $\mu M$ for different values of $\eta$.}
	\label{Fig: imag}
\end{figure}

In Fig.\ref{Fig: imag}, we compare the imaginary part of the frequency of the ground state (with $n=l=0$) for different values of $\eta$. The plot shows again that $|\omega_I|$ decreases with the increase of the black hole charge. We find that for $\mu M<0.25$ the logarithm of $|\omega_I|$ almost decreases uniformly as $\eta$ increases. This means that $|\omega_I|$ swiftly goes to zero in the limit $\eta\rightarrow\infty$.

\begin{figure*}
	\centering	
	\includegraphics[width=0.45\textwidth,height=0.35\textwidth]{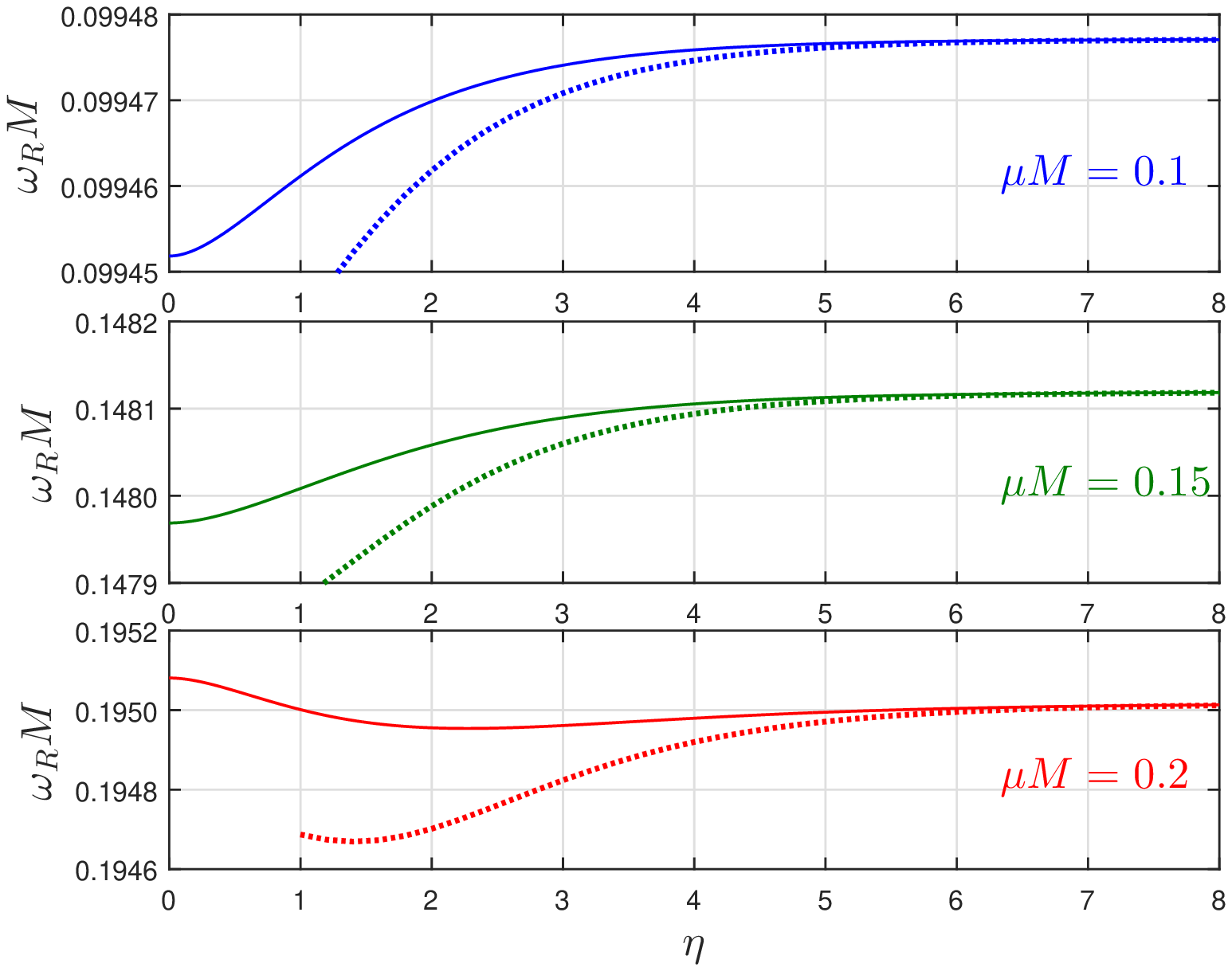}	
	\includegraphics[width=0.45\textwidth,height=0.35\textwidth]{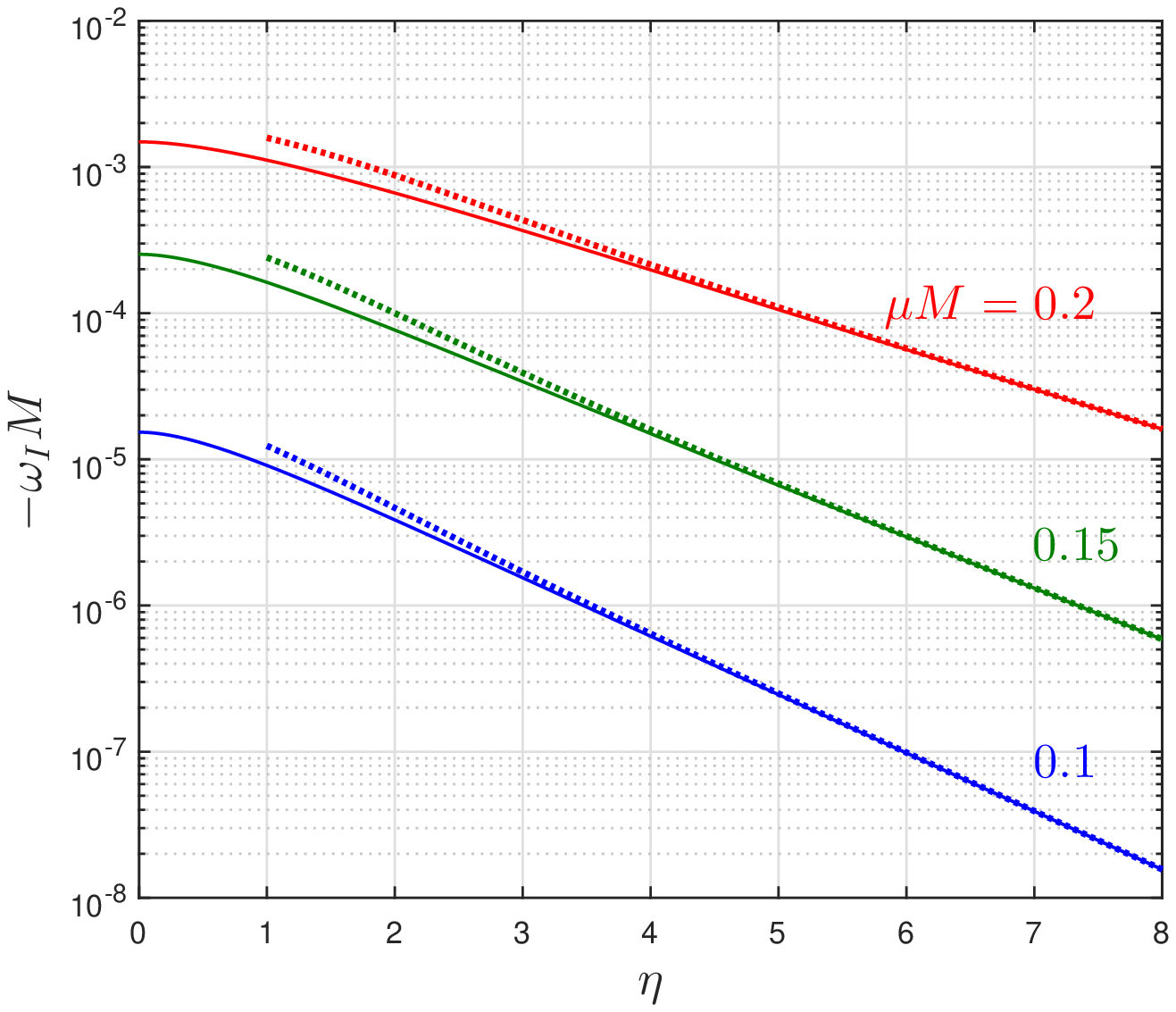}		
	\caption{The lowest quasibound state frequencies of $l=0$ state as functions of $\eta$, for $\mu M=0.1,0.15,0.2$. The left panel shows the real part $\omega_R$ of the frequency, and the right panel shows the imaginary part (or more precisely, $|\omega_I|$). In this plot, the solid lines denote the numerical results, while the dashed lines denote the analytical results obtained from Eq.(\ref{Eq: boundstate condition 2}).}
	\label{Fig: quasibound states num}
\end{figure*}

In Fig.\ref{Fig: quasibound states num}, we show the quasibound state frequencies of $l=0$ state as functions of $\eta$, for different values of $\mu M$. For both the real and imaginary parts of the frequency, numerical results agree quite well with the analytical ones, especially for large value of $\eta$. As expected, the imaginary part $|\omega_I|$ tends to zero exponentially.

\begin{figure}
	\centering	
	\includegraphics[width=0.45\textwidth,height=0.35\textwidth]{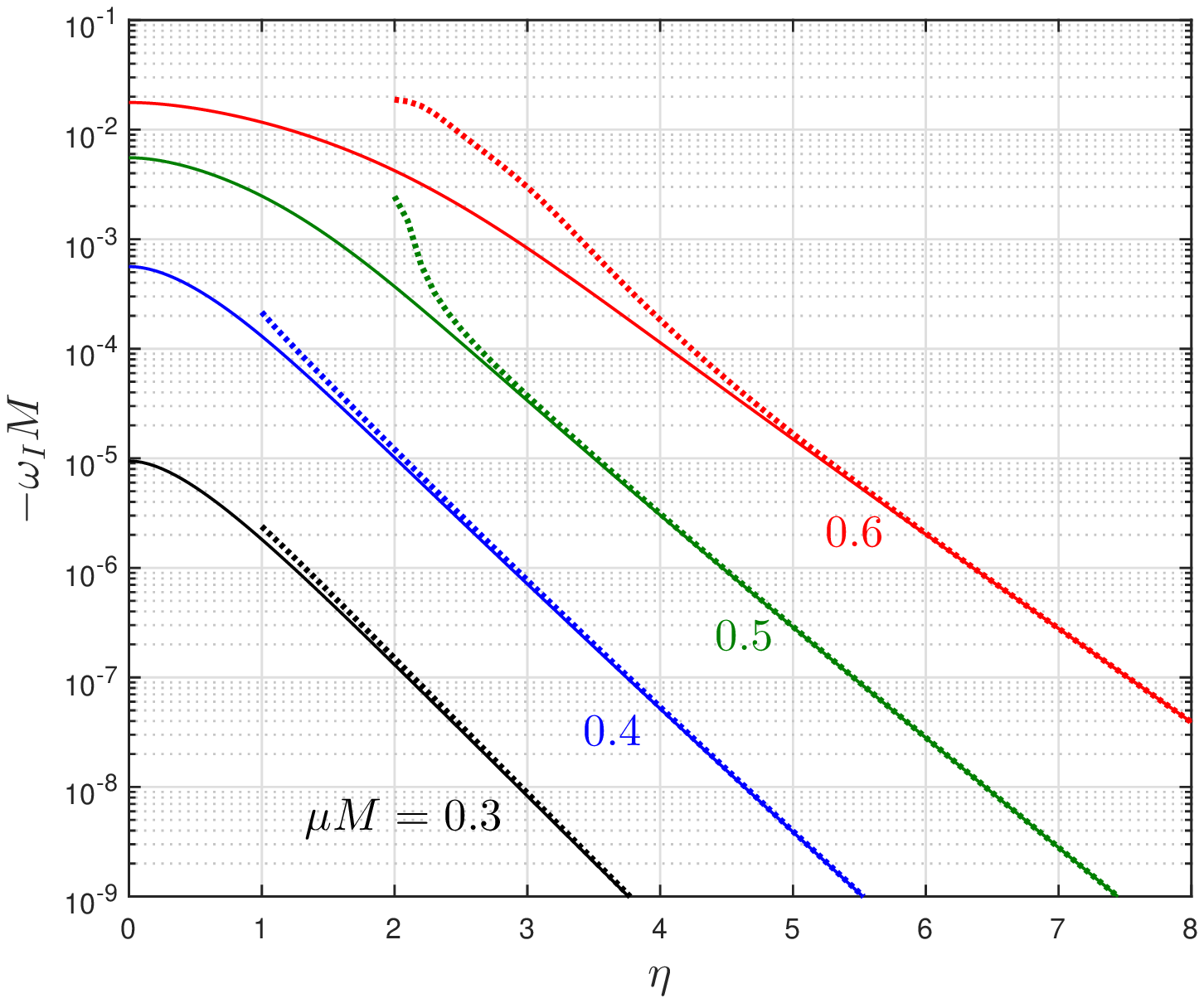}		
	\caption{The lowest quasibound state frequencies of $l=1$ state as functions of $\eta$, for $\mu M=0.3,0.4,0.5,0.6$. The solid lines denote the numerical results, while the dashed lines denote the analytical results obtained from Eq.(\ref{Eq: boundstate condition 2}).}
	\label{Fig: quasibound states l1}
\end{figure}

Figure \ref{Fig: quasibound states l1} compares the imaginary part of the frequencies of $l=1$ state for different values of $\mu M$. Again, we see that $|\omega_I|$ tends to zero exponentially at large $\eta$.
For lower mass coupling $\mu M$, $|\omega_I|$ goes down faster. A smaller coupling $\mu M$ implies a larger Compton wave length, which denotes a larger expectation of $r$ for the wave function. And a lower probability of tunnelling follows a smaller coupling $\mu M$, that is to say a smaller imaginary part of the eigen frequency.


\section{Discussion and conclusion}

  Bound state of hydrogen atom has fundamental importance, not only for quantization of matter, but also for quantization of electromagnetic field.
  Through transition between different energy levels, one directly demonstrates that the electromagnetic field is quantized. As it is well-known, the quantization
  of gravity is a long standing problem. A main reason why a full-fledged quantum gravity theory is still in absence is that we have no guide to develop such a theory from lab experiments
  or astrophysical observations. Similar to the case of  hydrogen atom, the transition between different energy levels of (quasi)bound state should emit or absorb graviton, i.e., a
  gravitational wave at a given frequency.  It is very difficult, if not impossible,  to detect the quantum property of gravitational waves from binary black holes \cite{DYSON:2013jra,Parikh:2020nrd,Hongsheng:2018ibg}. The transition between different
  energy levels of quasibound state of black hole leave finger print in the gravitational wave signals, if the progenitors of the wave have surrounding bounded articles.
  It may be beneficial to probe the quantum property of gravity field through analysis of gravitational wave from such progenitors.
  
  GMGHS black hole is a charged solution in dilatonic gravity. Dilatonic gravity is a minimal extension of general relativity, in which a new dilaton degree is introduced. GMGSH solution
  is not a hairy black hole, since the dilatonic charge is determined by the electromagnetic charge. Thus the dilatonic charge is not an independent new charge.

In this article, we study the quasibound state of massive scalar field in the GMGHS black hole spacetime.
We computed the eigen frequencies of quasbound state via both analytical and numerical method.
Results obtained from the two approaches are finely agree with each other.

We found that in the extremal limit $\eta\rightarrow\infty$ (or $q\rightarrow1$), the imaginary part of the frequency tends to zero exponentially.
This implies that massive scalar field configurations around a near extremal GMGHS black hole may be notably long-lived.
	
\begin{acknowledgments}
	We thank Dao-Jun Liu for helpful discussions. This work is supported by the National Key Research and Development Program of China (No. 2020YFC2201400), as well as Shandong Province Natural Science Foundation under grant No. ZR201709220395.
\end{acknowledgments}

\bibliography{boundstate}

\end{document}